\documentclass[]{IEEEtran}

\usepackage[pdftex]{graphicx}
\usepackage{amsmath}
\usepackage{amssymb}

\begin{document}

\title{Noise-Compensating Algebraic Reconstruction for\\ a Rotating Modulation Gamma-Ray Imager}

\author{B. Budden,
				G.~L. Case,
				M.~L. Cherry,~\IEEEmembership{Member,~IEEE}% <-this % stops a space
\thanks{All authors are with the Dept. of Physics \& Astronomy, Louisiana State University, Baton Rouge, LA 70803 (e-mail: bbudden@phys.lsu.edu).}}

%\author{Michael~Shell,~\IEEEmembership{Member,~IEEE,}
%        John~Doe,~\IEEEmembership{Fellow,~OSA,}
%        and~Jane~Doe,~\IEEEmembership{Life~Fellow,~IEEE}% <-this % stops a space
%\IEEEcompsocitemizethanks{\IEEEcompsocthanksitem M. Shell is with the Department
%of Electrical and Computer Engineering, Georgia Institute of Technology, Atlanta,
%GA, 30332.\protect\\
%% note need leading \protect in front of \\ to get a newline within \thanks as
%% \\ is fragile and will error, could use \hfil\break instead.
%E-mail: see http://www.michaelshell.org/contact.html
%\IEEEcompsocthanksitem J. Doe and J. Doe are with Anonymous University.}% <-this % stops a space
%\thanks{Manuscript received April 19, 2005; revised January 11, 2007.}}

% The paper headers
\markboth{IEEE Transactions on Nuclear Science}%
%\markboth{Journal of \LaTeX\ Class Files,~Vol.~6, No.~1, January~2007}%
{Shell \MakeLowercase{\textit{et al.}}: Bare Demo of IEEEtran.cls for Journals}

\IEEEcompsoctitleabstractindextext{%
\begin{abstract}
%\boldmath
A Rotating Modulator (RM) is one of a class of techniques for indirect imaging of an object scene by modulation and detection of incident photons. Comparison of the RM to more common imaging techniques, the Rotating Modulation Collimator and the coded aperture, reveals trade-offs in instrument weight and complexity, sensitivity, angular resolution, and image fidelity. In the case of a high-energy (hundreds of keV to MeV), wide field-of-view, satellite or balloon-borne astrophysical survey mission, the RM is shown to be an attractive option when coupled with a reconstruction algorithm that can simultaneously achieve super-resolution and suppress fluctuations arising from statistical noise. We describe the Noise-Compensating Algebraic Reconstruction (NCAR) algorithm, which is shown to perform better than traditional deconvolution techniques for most object scene distributions. Results from Monte Carlo simulations demonstrate that NCAR achieves super-resolution, can resolve multiple point sources and complex distributions, and manifests noise as fuzzy sidelobes about the true source location, rather than spurious peaks elsewhere in the image as seen with other techniques. 
\end{abstract}
% IEEEtran.cls defaults to using nonbold math in the Abstract.
% This preserves the distinction between vectors and scalars. However,
% if the journal you are submitting to favors bold math in the abstract,
% then you can use LaTeX's standard command \boldmath at the very start
% of the abstract to achieve this. Many IEEE journals frown on math
% in the abstract anyway. In particular, the Computer Society does
% not want either math or citations to appear in the abstract.

% Note that keywords are not normally used for peerreview papers.
\begin{IEEEkeywords}
Gamma-ray Astronomy, Image Reconstruction, Multiplexing, Scintillation Detectors.
\end{IEEEkeywords}}

% make the title area
\maketitle

\IEEEdisplaynotcompsoctitleabstractindextext
% \IEEEdisplaynotcompsoctitleabstractindextext has no effect when using
% compsoc under a non-conference mode.

% For peer review papers, you can put extra information on the cover
% page as needed:
% \ifCLASSOPTIONpeerreview
% \begin{center} \bfseries EDICS Category: 3-BBND \end{center}
% \fi
%
% For peerreview papers, this IEEEtran command inserts a page break and
% creates the second title. It will be ignored for other modes.
\IEEEpeerreviewmaketitle

%\IEEEdisplaynotcompsoctitleabstractindextext

%\IEEEpeerreviewmaketitle

% Introduction
\section{Introduction}

\IEEEPARstart{I}{maging} hard x-ray and gamma-ray photons (hundreds of keV to MeV) cannot be accomplished using focusing techniques, as with lower energy photons. A concise overview of several techniques is given in \cite{Caroli1987}. In one class of methods, incident photons are spatially or temporally modulated before detection. The recorded data are not a direct representation of the object scene, and so additional steps are required to deconvolve this information with a pre-determined instrument response function (i.e., system matrix).

A variety of deconvolution techniques have been developed across a wide range of applications, each employed for its demonstration of computational speed, noise suppression, resolving power, or fidelity; classes include statistical, algebraic, and ``ad-hoc'' algorithms. Algebraic techniques attempt to solve directly for the unknown image. Since this class requires convergence of the reconstruction to the data, noise in the data can be amplified and cause spurious peaks to arise. As we will show, however, an algebraic technique has a distinct advantage (the ability to achieve ``super-resolution'') that makes it entirely suitable for the rotating modulator, and with an appropriate non-linear step, statistical noise may be adequately suppressed.

% Rotational Modulation
\section{Rotating Modulation}
\label{sec:RM}

\begin{figure}
	\begin{center}
	\begin{tabular}{c}
		\includegraphics[width=2.3in]{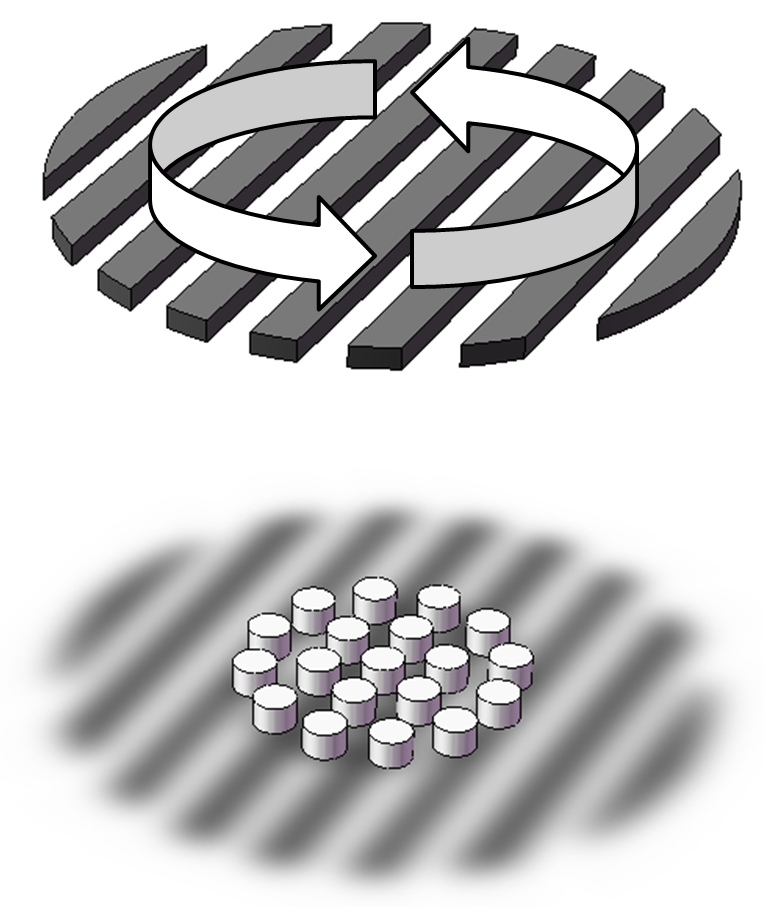} \\
		(a) Rotating Modulator \\
		\\
		\includegraphics[width=2.3in]{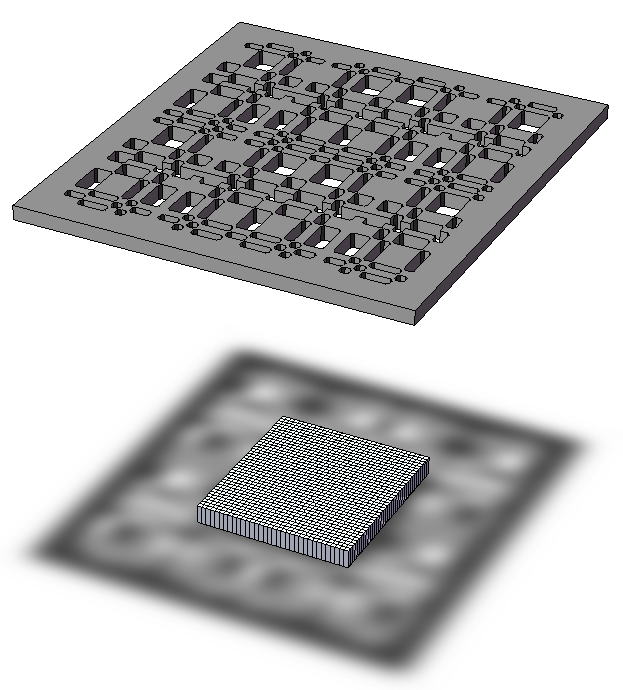} \\
		(b) Coded Aperture \\
		\\
		\includegraphics[width=2.3in]{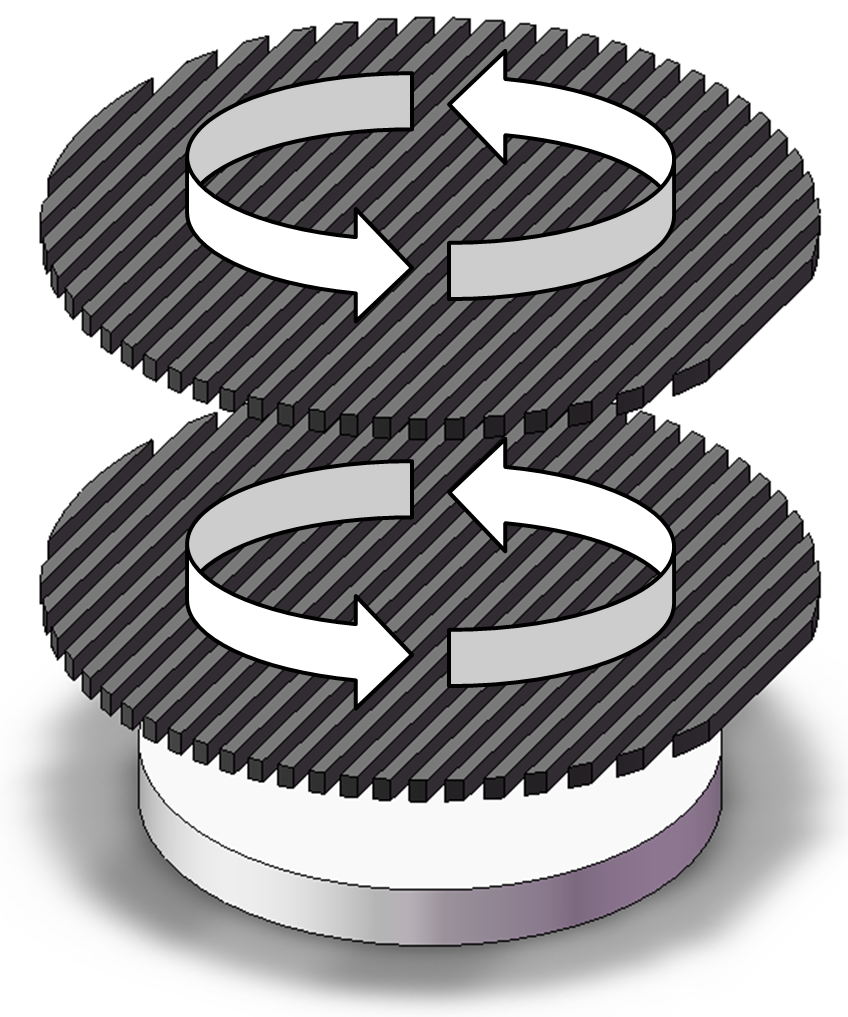} \\
		(c) Rotating Modulation Collimator
	\end{tabular}
	\end{center}
	\caption[example]
   { \label{fig:rm} 
   Basic structure of three multiplexing imagers. All instruments consist of a detection plane situated below a suspended mask which spatially or temporally modulates incident photons.}
\end{figure}

A Rotating Modulator (RM) \cite{Durouchoux1983, Dadurkevicius1985} is one of a class of temporal modulation imagers. It consists of a single grid of opaque slats spaced apart by equally-wide slits, suspended above a small array of circular non-imaging detectors (Fig. \ref{fig:rm}a). The detector diameters are equal to the slat/slit widths. The grid rotates, periodically blocking the transmission of incident photons from the object scene onto the detection plane. For each detector, this time-dependent shadow generates a characteristic time history of counts for the entire length of the exposure--varying from 0 to 100\% transmission--which is then folded modulo the rotational period of the grid. The profiles are combined with a collection of pre-calculated profiles to generate a cross-correlation image. Deconvolution of the cross-correlation with the instrument response produces the final image reconstruction.

To best understand the characteristics of the single-grid RM in the scope of the modulation class of imagers, it is useful to consider the better-known temporal imager, the two-grid Rotating Modulation Collimator (RMC) \cite{Mertz1967, Schnopper1968}, and the traditional spatial modulator for high-resolution gamma-ray imaging, the coded aperture \cite{Dicke1968, FenimoreCannon1978}. The latter incorporates a mask of opaque and transparent pixels (typically an equal number of each, resulting in a 50\% mask transmission), which spatially modulates incident photons onto a position-sensitive detection plane beneath (Fig. \ref{fig:rm}b). Angular resolution, $\delta\theta$, is defined by the ratio of mask pixel size, $a$, to mask/detector separation, $L$: $\delta\theta = a/L$. The detection plane should have spatial resolution half the mask pixel size or smaller (to satisfy the Nyquist condition) in order to resolve a shadow well enough to obtain the intrinsic resolution of the instrument. For a given detector pixel size, the maximum achievable resolution is limited due to a finite sampling of the shadow pattern. The resolving power may be increased only by further sub-dividing the detection plane. For a high-sensitivity, large-area instrument, this results in a high number of readout channels, increasing cost and complexity. (In the case of the CASTER \cite{McConnell2004} and EXIST \cite{Grindlay2001, Grindlay2009} designs for a satellite-based Black Hole Finder Probe gamma-ray telescope, the requirement for 10 minute-of-arc-scale angular resolution and meter-squared-scale sensitive area leads to in excess of $10^4$ -- $10^7$ readout channels.)

An RMC (Fig. \ref{fig:rm}c) consists of two offset concentric grids of equal-width slats and slits. One grid sits just above a single non-imaging detector, while the other grid is a distance $L$ overhead. As the two grids rotate in tandem, incident photons from off-axis sources within the field of view (FOV) will be periodically shadowed. A characteristic count profile is recorded by the detector, and then used for image reconstruction. As with the coded aperture, the geometric angular resolution is defined by the ratio of slit width to mask-detector separation, and so grids with small spacing can be constructed to provide excellent angular resolution. Due to the continuous nature of the measurement vector, the observation profiles may be sampled at higher frequency with no increased cost or complexity. With an appropriate reconstruction algorithm, resolution can be enhanced beyond the geometric resolution, although at lower imaging efficiency (see discussion in Sec. \ref{sec:sensitivity}). At high photon energies, the combination of large grid thickness and small slat spacing limits the maximum observable source angle due to mechanical collimation. With only a single detector, the RMC is a much simpler system than a coded aperture, though in practice, a single large-area detector would require multiple readout channels. The addition of a thick second grid adds to instrument weight and reduces the average transmission to 25\%. Furthermore, image reconstruction generates ``mirror'' sources, and the RMC is insensitive to sources directly overhead.

The RM may be considered a trade-off between the coded aperture and RMC, featuring temporal modulation and modest spatial resolution. It features a simple readout system, 50\% mask transmission, and relatively low weight. In analogy to the coded aperture and RMC, the intrinsic angular resolution is defined by the ratio of slit width to mask-detector separation. Since this width is set equal to the detector diameters in order to maximize sensitivity (by increasing the contrast in the count profiles), the resolution is effectively limited for a given mask-detector separation and detector diameter. Thus, for the RM to be a competitive alternative to other common multiplexing instruments, a reconstruction technique must be capable of resolving images beyond the intrinsic resolution, i.e. it must achieve ``super-resolution.'' The RM then compares favorably to a coded aperture with respect to its ability to sub-sample the observed data and go beyond the instrinsic resolving power and favorably to an RMC at high energies due to a wider achievable FOV and the weight of only a single mask.

While the RMC has seen limited use in gamma-ray imaging,--e.g. Ariel-V \cite{Carpenter1976} (3 - 7 keV), SAS-3 \cite{Doxsey1976} (1.5 - 60 keV), RHESSI \cite{Hurford2002} (3 keV - 17 MeV)--the authors are aware of no RM previously used in astronomy. The WATCH experiment \cite{Lund1981} was a self-described single-grid RMC, although its design was perhaps closer to that of an RM with some key differences (including slat detectors instead of circular detectors). Other than a prototype RM recently constructed in our laboratory at Louisiana State University (LSU) \cite{BuddenIEEE2008}, we are aware of only one other RM in development \cite{Shih2008}. Our intention has been to demonstrate the feasibility of the RM design as a solution to gamma-ray imaging with wide FOV and good sensitivity and angular resolution. This requires the ability of an image reconstruction algorithm to resolve beyond the instrinsic resolution limit of the instrument. We will show that a novel reconstruction algorithm based on an algebraic solution can produce images with resolution several times better than that defined by the instrument geometry. Furthermore, multiple and complex source distributions can be well represented, no object information is required a priori, and background and counting noise can be adequately suppressed.

\section{The Imaging Problem}

The data observed by a multiplexing instrument, $O(m)$, may be defined by the counts observed in each data bin $m$:
\begin{equation}
	O(m) = \sum_n P(m,n) S(n) + B(m) + N(m),
	\label{eq:o}
\end{equation}
where $S(n)$ is the object scene, $B(m)$ is a background offset, $N(m)$ is the measured noise from scene and background combined, and $P(m,n)$ is the instrument response, a function that transforms the scene information from image space, $n$, to data space, $m$. In practice, a background subtraction removes the $B(m)$ term (though the background \emph{noise} remains), and so it is absent from the formulas henceforth. In matrix form, Eq. \ref{eq:o} is written more succinctly as
\begin{equation}
	O_m = P_{mn}S_n + N_m.
	\label{eq:oMatrix}
\end{equation}

Solving this equation for the object scene $S_n$ (i.e. deconvolving $O_m$ and $N_m$ with the instrument response $P_{mn}$) is the goal of image reconstruction. If $P_{mn}$ were non-singular, then the object scene could be written
\begin{equation}
	S_n = P^{-1}_{nm} (O_m - N_m).
	\label{eq:sWritten}
\end{equation}
In general, however, $P$ is neither square nor has an inverse. Even if $P$ were non-singular, numerical uncertainties make this method impossible in practice. For these reasons, a precursor to most techniques is to pre-condition the matrix $P$ by multiplying both sides of Eq. \ref{eq:oMatrix} by the transpose of the instrument response:
\begin{equation}
	P^T_{n'm} O_m = P^T_{n'm} ( P_{mn} S_n + N_m),
\end{equation}
or more simply,
\begin{equation}
	C_{n'} = P'_{n'n} S_n + N'_{n'},
	\label{eq:c}
\end{equation}
where the cross-correlation image is $C_{n'} = P^T_{n'm} O_m$, and the point-spread function (PSF) is $P'_{n'n} = P^T_{n'm} P_{mn}$. By definition, $P'$ is square-symmetric and will tend towards being diagonally-dominant for local PSFs that resemble a two-dimensional delta function. For this reason, Eq. \ref{eq:c} is a more suitable system to solve than Eq. \ref{eq:oMatrix}. The unknown noise term, $N_{n'} = P^T_{n'm} N_{m}$, complicates the solution. The deconvolution technique employed must therefore compensate and correct for noisy data, while reconstructing true sources.

% Technique
\section{RM Instrument Response}
\label{sec:insresp}

\begin{figure}
	\begin{center}
	\includegraphics[width=3.0in]{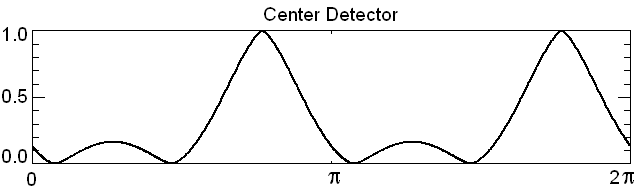} \\
	\includegraphics[width=3.0in]{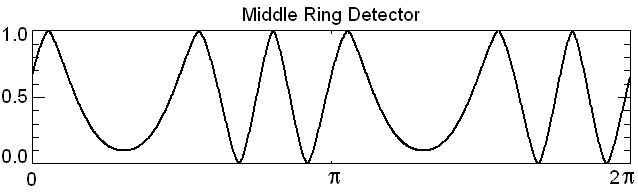} \\
	\includegraphics[width=3.0in]{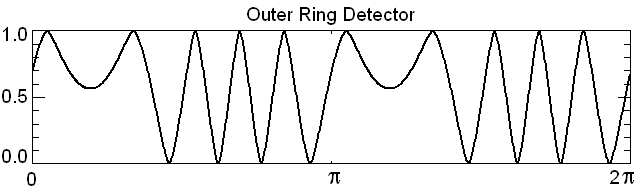} \\
	\end{center}
	\caption[example]
   { \label{fig:profiles} 
   Deterministic count profiles from three separate detector locations for a single point source. The horizontal scale is grid rotation phase, while the vertical scale is unit source intensity.}
\end{figure}

As the RM grid rotates, each detector $d$ produces a time history of counts,
\begin{equation}
	O_d(t) = \sum_n P_d(t,n)S(n) + N_d(t).
	\label{eq:oRM}
\end{equation}
The instrument response $P_d(t,n)$ is a collection of pre-calculated count profiles for all possible $n$ source locations within the object scene. These profiles are determined analytically for an RM with slat width $a$ by \cite{Dadurkevicius1985}
\begin{multline}
	P_d(t,n) = \\ 
	1 - \gamma F\left( 1 - \left| \left| \frac{r(n)}{a} \cos{(\xi(t)+\xi_0)} \right| \mbox{ mod } 2 - 1 \right| \right),
	\label{eq:p}
\end{multline}
where $\xi(t)$ is the angular orientation of the grid at time $t$. The parameter $r(n)$ is the distance in the plane of detection between the detector and the axis of the projected grid shadow for a source located at scene location $n$. If the RM detectors are stationary, $r(n)$ remains constant with time. For a grid height $L$, this distance is defined in terms of the source azimuth $\phi$ and zenith $\theta$ by
\begin{equation}
	r(n) = \sqrt{ (x_0 + L \tan\theta \cos\phi)^2 + (y_0 + L\tan\theta\sin\phi)^2 },
	\label{eq:r}
\end{equation}
where $x_0$ and $y_0$ are the detector position. The instrument is assumed to be fixed relative to the sky; however, if the instrument moves with a known behavior, a conversion can be made to move these coordinates to a global reference frame. (Such an operation would be necessary for a balloon-borne or satellite instrument.) The function $F(\tau)$ describes the percentage of circular detector area covered by a shadow that has moved a fraction of the total diameter, $\tau$, across its face:
\begin{equation}
	F(\tau) = \frac{1}{\pi} \cos^{-1}(1-2\tau) - \frac{2}{\pi}(1-2\tau)(\tau-\tau^2)^{1/2}.
	\label{eq:F}
\end{equation}
The coefficient $\gamma$ accounts for transmission through the mask and is dependent on the mask thickness $h$, density $\rho$, and mass absorption coefficient $\sigma$ for a particular photon energy, as well as the angle of photon incidence (i.e. zenith):
\begin{equation}
	\gamma = 1-\exp{\left[ - \frac{ h \rho \sigma}{\cos\theta} \right]}.
	\label{eq:gamma}
\end{equation}
It is important to note that Eq. \ref{eq:p} is an approximation only; it ignores photons that ``clip'' the slat edges. (Description of an advanced characteristic formula to extend these results by including edge clipping effects will be presented separately.) Fig. \ref{fig:profiles} shows count profile examples of the same source viewed by detectors at three different positions.

\begin{figure}
	\begin{center}
		\includegraphics[width=3.0in]{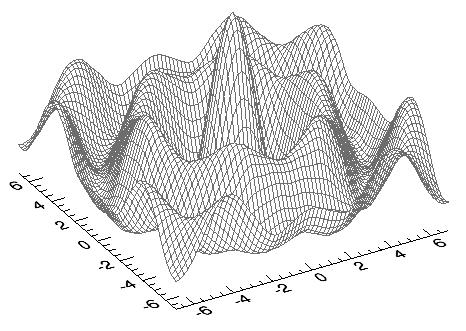}
	\end{center}
	\caption[example]
   { \label{fig:c} 
   Cross-correlation image surface mesh of a single point source centered in the FOV. The concentric ``ring'' nature of the PSF is apparent. (Axes in degrees)}
\end{figure}

The transformation is made from the temporal observation to image space by a cross-correlation of the data with the transpose of the instrument response summed over the detectors,
\begin{equation}
	C(n) = \sum_d \sum_t P_d(t,n)O_d(t).
	\label{c_full}
\end{equation}
Substituting $O(t)$ from Eq. \ref{eq:oRM} into Eq. \ref{c_full}, we see that the cross-correlation image is the object scene correlated with a single PSF:
\begin{equation}
	C(n) = \sum_{n'} P'(n',n) S(n') + N(n),
	\label{eq:cRM}
\end{equation}
where
\begin{equation}
	P'(n',n) = \sum_d\sum_t P_d(t,n') P_d(t,n).
	\label{eq:pRM}
\end{equation}
Conceptually, an element of $P'(n',n)$ may be thought of as the relative brightness of pixel $n'$ for a point source located at location $n$ (or vice-versa, due to symmetry). Eqs. \ref{eq:cRM} and \ref{eq:pRM} are analogous to the dirty map and dirty beam, respectively, in radio astronomy. For the RM, the PSF appears as a central peak with concentric ``ring'' sidelobes (Fig. \ref{fig:c}), which may be described mathematically by a zeroth order Bessel function \cite{Mertz1976}.

% RM Sensitivity
\section{RM Sensitivity}
\label{sec:sensitivity}

While a detailed examination of RM sensitivity is beyond the scope of this paper, it is necessary to present the formula for the signal-to-noise ratio (SNR) of a reconstructed image to compare the three discussed multiplexing imagers, particularly for the case when super-resolution is achieved. For an RM, an RMC, or a coded aperture, the statistical significance of an observation at location $n$ of the raw image (Eq. \ref{eq:c}) is found by dividing the value in that element by the standard deviation in that element \cite{Durouchoux1983}. For a total source rate $S(n)$ as measured by all detectors in the absence of modulation and total background rate $B$, the SNR is approximated for the case where $B \gg S(n)$ as
\begin{equation}
	\mbox{SNR}(n) \approx S(n) \left(\frac{T}{B}\right)^{\frac{1}{2}} \frac{1}{\eta^2} \left( \overline{P(n)^2} - \overline{P(n)}^2\right)^{\frac{1}{2}},
	\label{eq:SNR}
\end{equation}
where $T$ is the source exposure time, and $\eta$ describes the increased sub-sampling of the object scene beyond the geometric resolution $\delta\theta$. Each factor $\eta$ of super-resolution will therefore result in a decrease in the instrument efficiency of $\eta^{-2}$. According to Eq. \ref{eq:SNR}, the SNR of a multiplexing instrument observation is related to a non-imaging instrument by the square root of the variance of the instrument response. (In units of $S(T/B)^{1/2}$, the SNR of an observation by a non-imaging instrument is 1.)

For a coded aperture, the mask pattern is composed of ones and zeroes, which represent open and closed elements, respectively. Coded patterns are specifically chosen with standard deviation 0.5 for all $n$, the maximum deviation possible with an instrument that blocks 50\% of the incident photons. RMC sensitivity can likewise be calculated, but due to lower throughput and non-ideal instrument response, the average deviation (and thus statistical significance) is approximately 0.15. Unlike the coded aperture or non-imaging instrument, the sensitivity for an RMC is not uniform across the sky, and falls to zero directly overhead. (The use of multiple RMCs with varying pointing directions will, however, smooth out sensitivity across the object scene.)

\begin{figure}
	\begin{center}
		\includegraphics[width=3.2in]{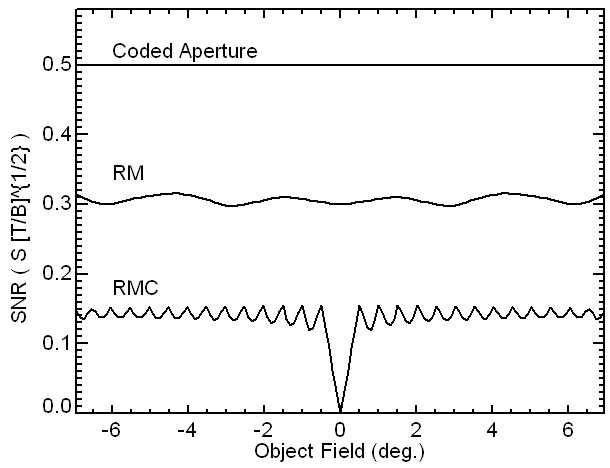}
	\end{center}
	\caption[example]
   { \label{fig:sensitivity} 
   Sensitivity plot for three multiplexing imagers, relative to $S\sqrt{T/B}$.}
\end{figure}

Due to the fact that the RM incorporates multiple detectors, each with its own instrument response, the statistical significance of an observation must include the variance of each detector's response:
\begin{equation}
	\mbox{SNR}_{RM}(n) \approx \frac{S(n)}{\eta^2} \left( \frac{T}{BD} \sum_d \left[ \overline{P_d(n)^2} - \overline{P_d(n)}^2 \right] \right)^{\frac{1}{2}},
	\label{eq:SNR_RM}
\end{equation}
where $D$ is the total number of detectors. (The requirement that slat/slit width be equal to the detector diameter can be determined from this formula; unequal widths would decrease the profile contrast, inherently lowering the variance and thus the sensitivity.) Due to the addition of multiple response variances, the resulting sky sensitivity is more uniform, and due to the higher throughput, inherently better (approximately 0.31) than the RMC. Because the PSF is distributed over a central peak and broad sidelobes (Fig. \ref{fig:c}), the RM sensitivity is, however, less than that of the coded aperture. Figure \ref{fig:sensitivity} shows the sensitivity curves for typical configurations of these three multiplexing instruments.

% Reconstruction Technique
\section{Reconstruction Technique}

The square-symmetric nature of $P'_{n'n}$ makes the system suitable for a wide variety of deconvolution techniques to solve for $S$, which have been developed across a broad spectrum of applications. For example, CLEAN \cite{Hogbom1974} and the Maximum Entropy Method (MEM) \cite{Jaynes1957} are widely used in radio interferometry, while the Algebraic Reconstruction Technique (ART) \cite{Gordon1970} and Maximum Likelihood Expectation Maximization (MLEM) \cite{Dempster1977} are commonly implemented in medical imaging. These traditional reconstruction techniques are unsuitable for the RM. CLEAN does not perform well when many sources are present due to the extended nature of the RM PSF, and it is unable to provide super-resolution. While we have previously shown imaging results using MEM \cite{BuddenIEEE2008}, it is also unable to provide super-resolution or resolve multiple and complex sources. ART is subject to noise fluctuations due to its requirement that the reconstruction agrees exactly with the data. MLEM produces good results with super-resolution and suppression of noise, but is slow; the MLEM algorithm assumes that the data are Poisson distributed, and so deconvolution must take place directly from the temporal domain, where detector data can not be combined to speed up reconstruction.

We have developed a new technique derived from Direct Demodulation (DDM) \cite{Li1994}. In simulations, DDM has been previously demonstrated to provide super-resolution with RMCs \cite{chen1998} and we have shown \cite{BuddenSPIE2009} the ability of DDM with an RM to achieve $\sim 8\times$ the geometric resolution and to resolve multiple and complex sources when little or no noise exists. For the reconstruction of actual measured data in the presence of background, we have modified the DDM routine and implemented a key step to compensate for noisy data. We refer to this technique as Noise-Compensating Algebraic Reconstruction (NCAR).

The Gauss-Seidel iterative method is used to solve algebraically for the object scene. The $(k+1)$ iteration of the reconstruction, $f(n)$, is determined by \cite{Barrett1994}
\begin{multline}
	f^{(k+1)}(n) = \frac{1}{P'(n,n)} \left( C(n) - \sum_{m<n} P'(m,n) f^{(k+1)}(m) \right. \\
 \left. - \sum_{m>n} P'(m,n) f^k(m) \right),
\label{eq:gs}
\end{multline}
with $f^0(n) = 0$. Gauss-Seidel uses the results of calculations from the same iteration, and so to prevent pixel bias, it is prudent to randomize the order of $m$ every iteration. A positivity condition is enforced as a physical constraint on each pixel of $f^{(k+1)}(n)$.

\begin{figure}
	\begin{center}
	\begin{tabular}{c}
		\includegraphics[width=2in]{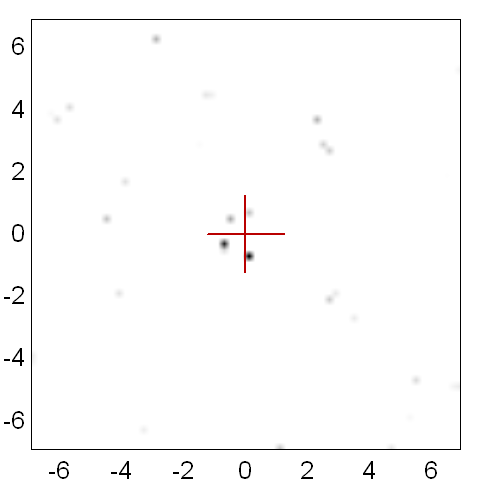} \\
		\includegraphics[width=3in]{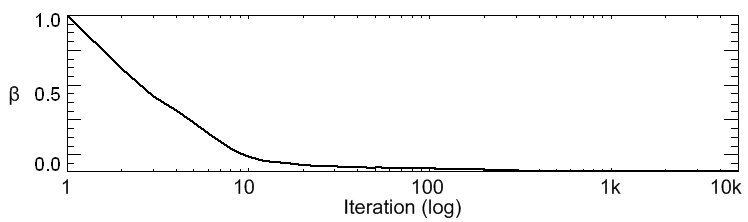} \\
		(a) \\
	 	{  }\\
		\includegraphics[width=2in]{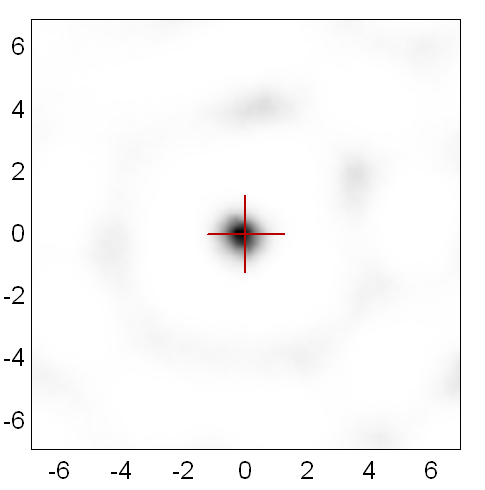} \\
		 \includegraphics[width=3in]{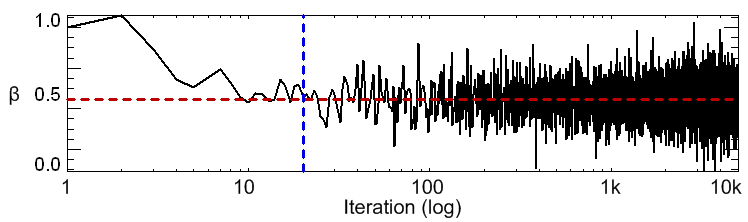} \\
		(b) \\
	\end{tabular}
	\end{center}
	\caption[example]
   { \label{fig:agreement} 
   Simulation of a single point source centered in the FOV, showing the reconstructed image (with actual source location designated by a $+$ symbol) and corresponding normalized residual summation, $\beta$, plotted against iteration number using (a) Direct Demodulation, and (b) with the noise compensation additive, $R[\sigma(n)]$, given in Eq. \ref{cWithR}. In (b), the vertical dashed line corresponds to $\kappa$, the iteration at which the maximum agreement has been reached, and the horizontal dashed line corresponds to the residual convergence value given the noisy data. Note the contrast to (a), which converges indefinitely to zero (i.e. perfect agreement).}
\end{figure}

A normalized parameter $\beta$ may be defined, which sums over all pixels in the residual to examine the agreement of the reconstruction with the cross-correlation,
\begin{equation}
	\beta^{(k+1)} \equiv \frac{1}{\beta^{(0)}} \sum_{n'} \left| \sum_n P'(n',n)f^{(k+1)}(n) - C(n') \right|.
	\label{f:agreement}
\end{equation}
In Fig. \ref{fig:agreement}a, this residual summation is plotted against the iteration number for a sample reconstruction. The successive reconstructions converge to the data (i.e. $\beta$ shrinks) for an indefinite number of iterations. This agreement, however, includes any fluctuations from noise within the data, and so this method causes spurious peaks and poor location accuracy of the true source, as seen in the reconstruction, when the signal-to-noise ratio (SNR) is low.

This ``noise amplification'' is the primary problem in image reconstruction using an algebraic technique. Regularization methods (e.g. smoothing, pixel-to-pixel variation constraints) may be employed, but this regularized image is still forced to agreement with the data, $C(n)$. The NCAR technique allows for deviation from the noisy data, by replacing $C(n)$ in Eq. \ref{eq:gs} with $C^{(k+1)}(n)$, such that
\begin{equation}
	C^{(k+1)}(n) \equiv C(n) + R^{(k+1)}[\sigma(n)].
	\label{cWithR}
\end{equation}
The function $R[\sigma]$ provides a random number from a normal distribution with average zero and standard deviation $\sigma$, where $\sigma(n)$ is the error in each pixel of the cross-correlation. If the error is assumed to be only from Poisson uncertainty, it can be shown that
\begin{equation}
	\sigma(n) = \sqrt{\sum_m O(m) P(m,n)^2}.
	\label{eq:sigma}
\end{equation}

The reconstruction and residual summation, $\beta$, are plotted for NCAR in Fig. \ref{fig:agreement}b. The agreement between reconstruction and data fluctuates about an envelope showing initial convergence followed by a leveling off beginning at some iteration $\kappa$. Once this minimum $\beta$ is reached, the image has converged as well as possible to the data given the noise. Each successive iteration still contains spurious peaks arising from noise fluctuations, but because the component $R[\sigma(n)]$ has been added to randomize the noise, the amplitude and location of these peaks are unique to each iteration, while the iterative images share the true sources. By now averaging over all the images after iteration $\kappa$, the spurious random peaks are suppressed while the true source distributions remain:
\begin{equation}
	F^{(k+1)}(n) = \frac{f^{(k+1)}(n) + (k-\kappa)F^{k}(n)}{k+1-\kappa}, \quad k+1 > \kappa.
\end{equation}
$F^{(k+1)}(n)$ is the final reconstruction shown in Fig. \ref{fig:agreement}b. Using NCAR, statistical uncertainty in the data is manifest as an increase in the size of the reconstructed source in the form of fuzzy sidelobes. In cases where the actual size of the source is desired, NCAR may not be desirable; but with large noise in the data, an algebraic technique alone will produce spurious peaks and may not represent the true source size anyway. When it is acceptable that source size in the reconstructed image relates to uncertainty in the data, however, NCAR produces image reconstructions with a significant reduction in noise fluctuations compared to results using MEM, CLEAN, or algebraic solutions, as will be demonstrated in detail in Sec. \ref{sec:results}.

%Simulation
\section{Simulation}
\label{sec:sim}

Monte Carlo simulations were performed to test and verify the NCAR technique. The software was custom written for this project in IDL. The instrument response and cross-correlation images are calculated as described in Sec. \ref{sec:insresp}. Since the instrument response assumes far-field imaging, sources are simulated at an ``infinite'' distance from the detector; i.e. photons from the same source arrive at equal angles of incidence.

\begin{figure}
	\begin{center}
		\includegraphics[width=3in]{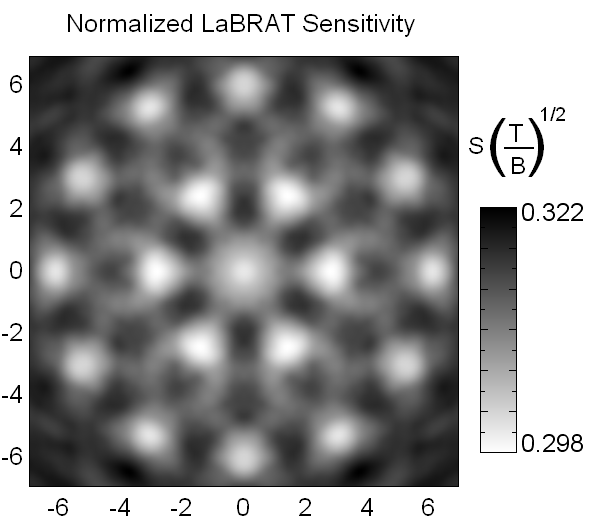}
	\end{center}
	\caption[example]
   { \label{fig:labrat_sens} 
   Contour plot of the relative sensitivity of LaBRAT's FOV, with the parameters used for the simulations described in Sections \ref{sec:sim} and \ref{sec:results}. Values are normalized to $S(T/B)^{1/2}$. Axes are in degrees.}
\end{figure}

The geometry of the simulated RM reproduces that of a laboratory prototype, the Lanthanum Bromide-based Rotating Aperture Telescope (LaBRAT), developed at LSU \cite{BuddenIEEE2008}. LaBRAT's detection plane is composed of nineteen cylindrical Cerium-doped Lanthanum Bromide (Ce:LaBr$_3$) scintillators, each 3.8 cm diameter $\times$ 2.5 cm thick. They are positioned in a hexagonal layout and coupled to 3.8 cm PMTs. The mask is a grid of eight 3.8 cm $\times$ 61 cm $\times$ 1.9 cm thick lead slats spaced 3.8 cm apart and sitting $\sim 1.2$ m away from the detectors. This arrangement defines a 13.8$^{\circ}$ diameter FOV and a 1.9$^{\circ}$ FWHM geometric angular resolution. As a result of the 19 detectors used in LaBRAT, the sensitivity is fairly uniform. As seen in Fig. \ref{fig:labrat_sens}, where white indicates regions of the object scene that are unmodulated from the viewpoint of one of the detectors (i.e. has less imaging sensitivity), the overall relative sensitivity is close to 0.31 over the entire scene. The pattern in the figure thus mimics the detector layout. Additional dithering of the telescope orientation about the vertical (as would occur in a balloon-borne instrument) would result in additional uniformity across the FOV.

\begin{figure}
	\begin{center}
	\begin{tabular}{@{}c@{}c@{}}
		Object Scene & Raw Image \\
		\includegraphics[width=1.73in]{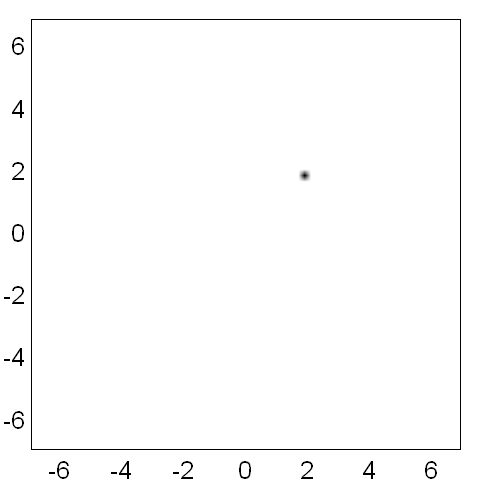} & \includegraphics[width=1.73in]{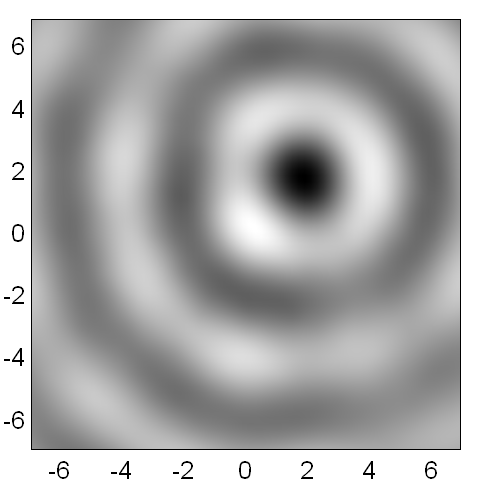} \\
		MEM & CLEAN \\
		\includegraphics[width=1.73in]{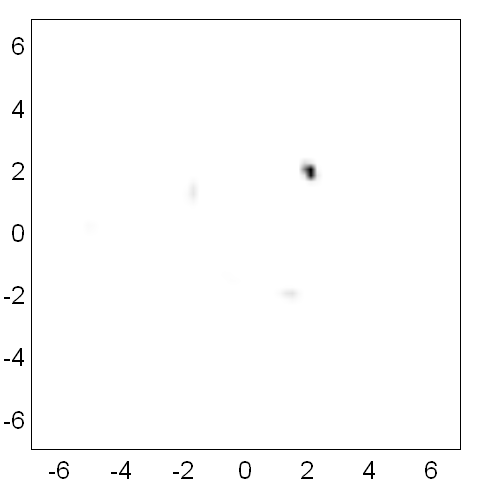} & \includegraphics[width=1.73in]{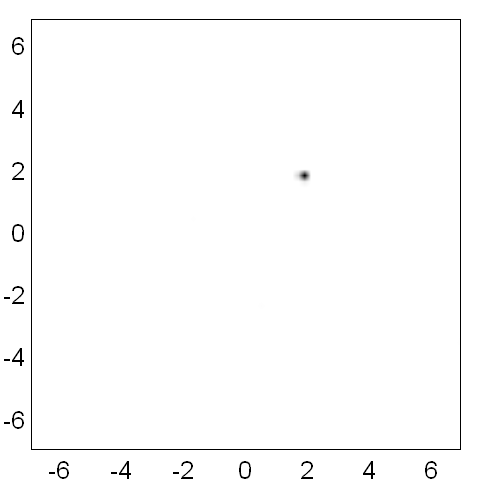} \\
		DDM & NCAR \\
		\includegraphics[width=1.73in]{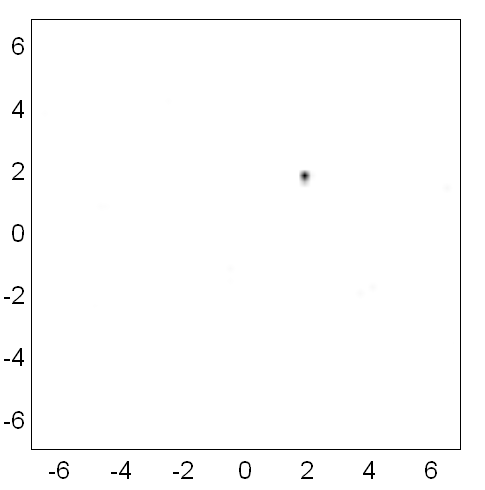} & \includegraphics[width=1.73in]{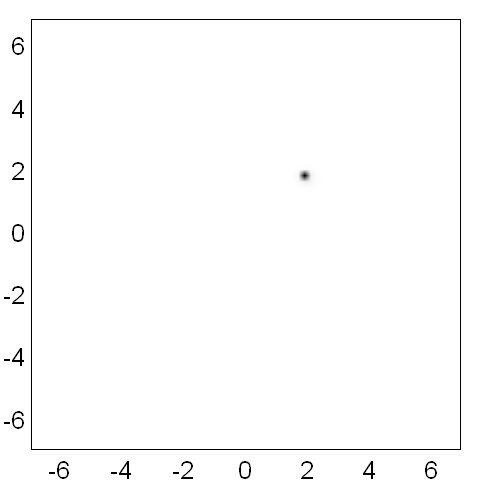} \\
		\end{tabular}
	\end{center}
	\caption[example]
   { \label{fig:sim_results_a2} 
   Monte Carlo Simulation results for a single point source and no background. Object scene is shown at upper left, raw cross-correlation image (Eq. \ref{eq:cRM}) is at upper right, and results of MEM, CLEAN, DDM and NCAR are shown below. All reconstruction techniques perform reasonably well. (Axes in degrees) \\ { }}
\end{figure}

\begin{figure}
	\begin{center}
	\begin{tabular}{@{}c@{}c@{}}
		Object Scene & Raw Image \\
		\includegraphics[width=1.73in]{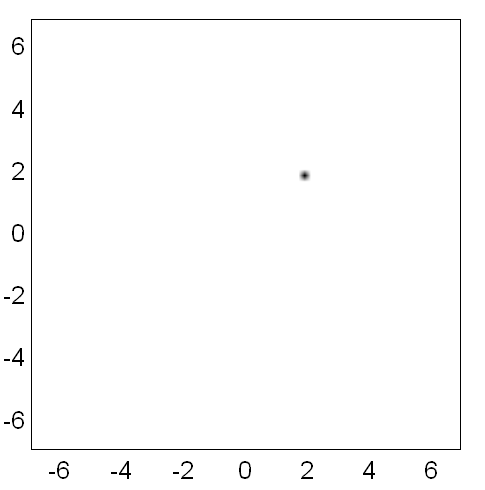} & \includegraphics[width=1.73in]{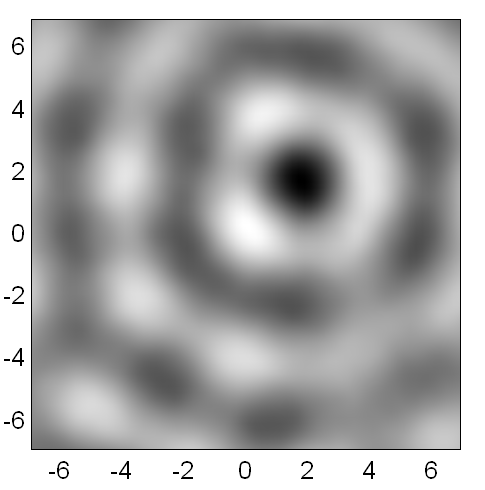} \\
		MEM & CLEAN \\
		\includegraphics[width=1.73in]{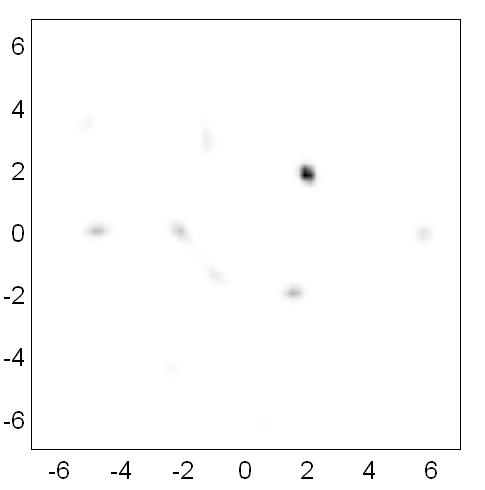} & \includegraphics[width=1.73in]{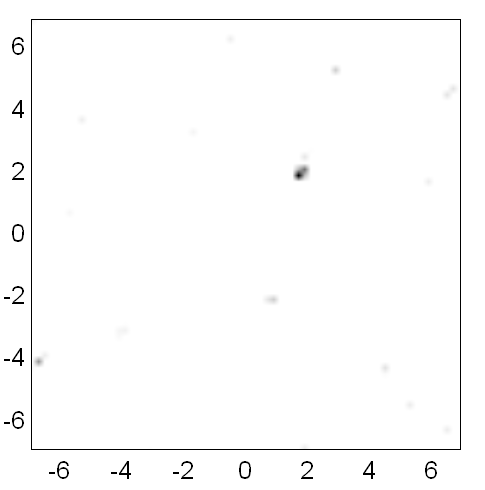} \\
		DDM & NCAR \\
		\includegraphics[width=1.73in]{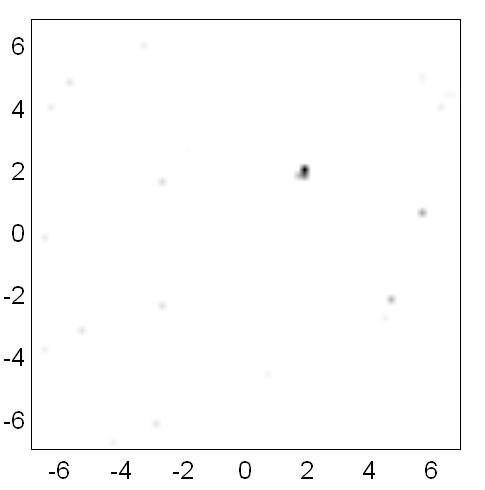} & \includegraphics[width=1.73in]{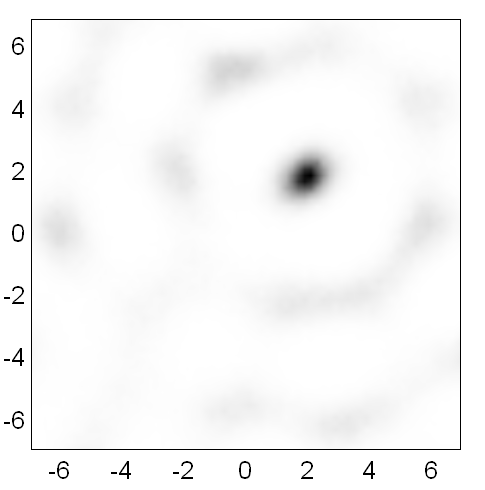} \\
		\end{tabular}
	\end{center}
	\caption[example]
   { \label{fig:sim_results_a} 
   Monte Carlo Simulation results for a single point source with background. Note that, in the presence of background, the MEM, CLEAN, and DDM reconstructions contain spurious point-like sources and the image does not convey the uncertainty of the location of the true source. NCAR, however, smooths over the noise contributions, and places an uncertainty on the true source that is related to the SNR of the measurement. (Axes in degrees)}
\end{figure}

\begin{figure}
	\begin{center}
	\begin{tabular}{@{}c@{}c@{}}
		Object Scene & Raw Image \\
		\includegraphics[width=1.73in]{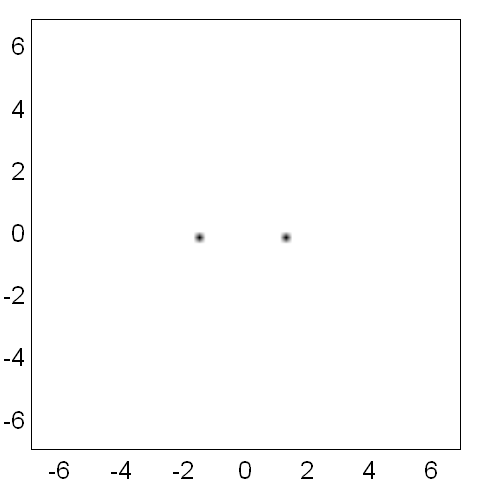} & \includegraphics[width=1.73in]{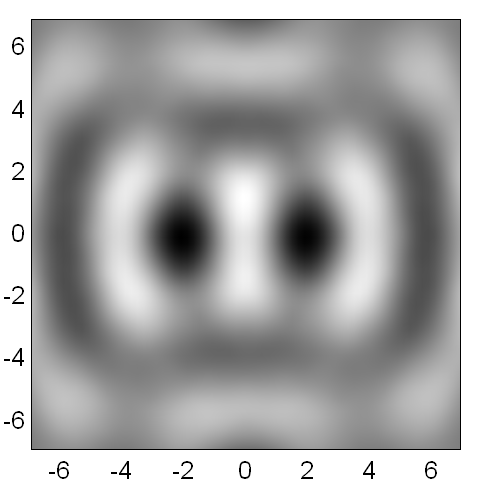} \\
		MEM & CLEAN \\
		\includegraphics[width=1.73in]{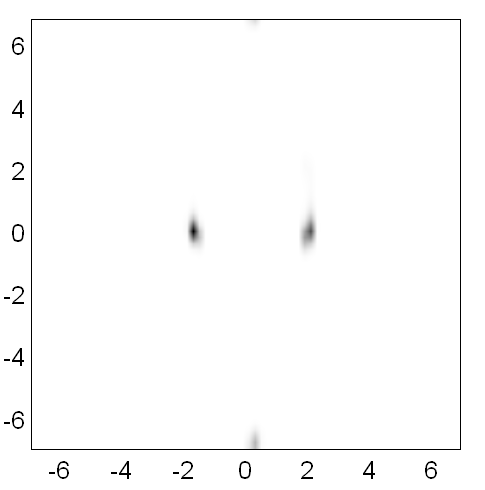} & \includegraphics[width=1.73in]{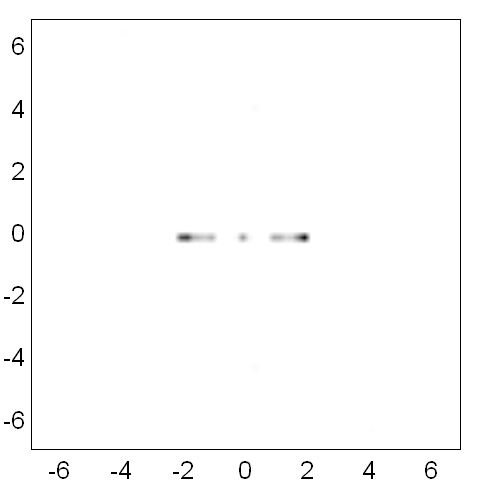} \\
		DDM & NCAR \\
		\includegraphics[width=1.73in]{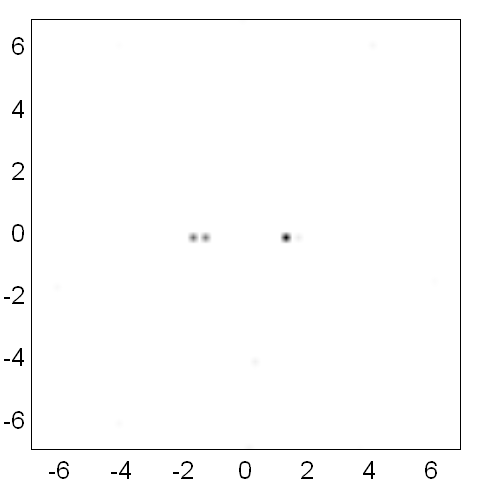} & \includegraphics[width=1.73in]{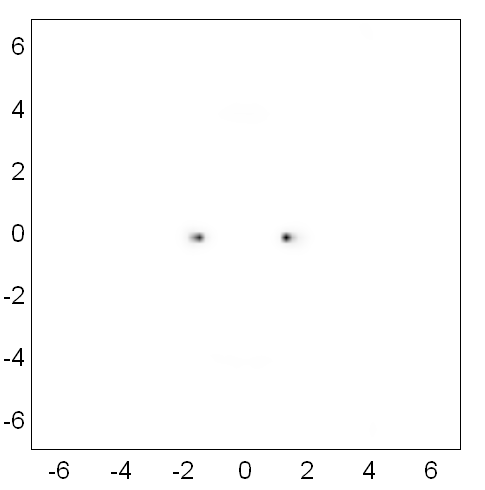} \\
		\end{tabular}
	\end{center}
	\caption[example]
   { \label{fig:sim_results_b} 
   Monte Carlo Simulation results for two equal intensity sources separated by 3$^{\circ}$. Statistical and algebraic reconstruction techniques are both capable of resolving two sources that are at an angular separation greater than the geometric resolution defined by the instrument. (Axes in degrees)}
\end{figure}
		
\begin{figure}
	\begin{center}
	\begin{tabular}{@{}c@{}c@{}}
		Object Scene & Raw Image \\
		\includegraphics[width=1.73in]{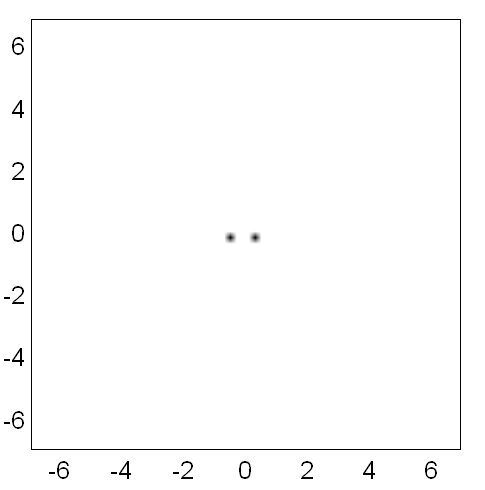} & \includegraphics[width=1.73in]{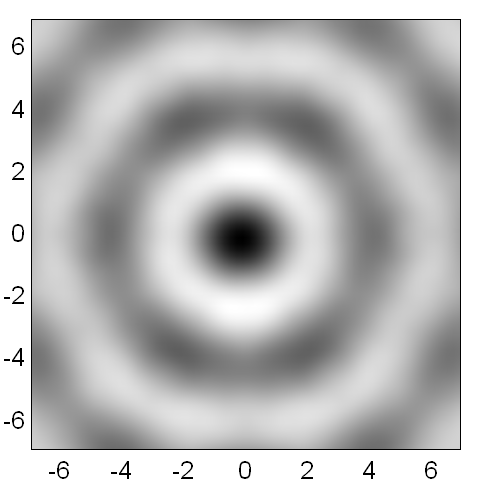} \\
		MEM & CLEAN \\
		\includegraphics[width=1.73in]{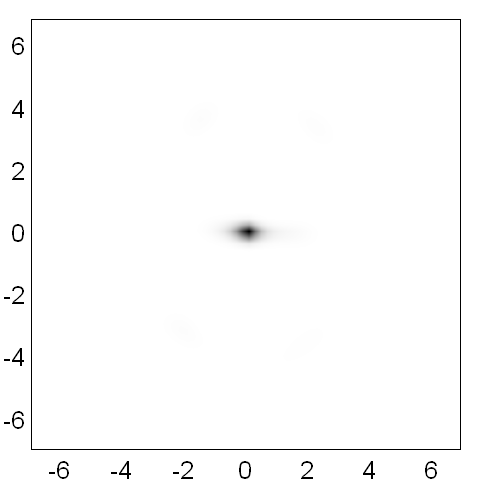} & \includegraphics[width=1.73in]{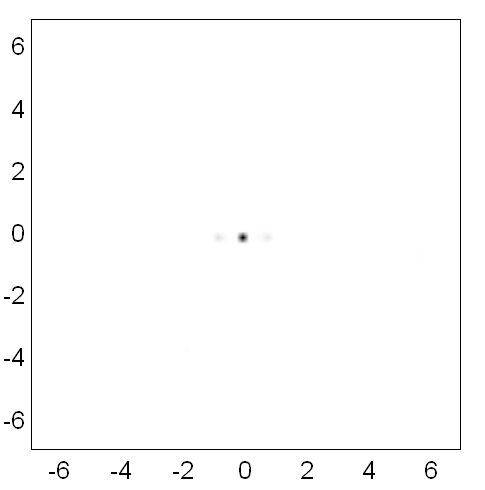} \\
		DDM & NCAR \\
		\includegraphics[width=1.73in]{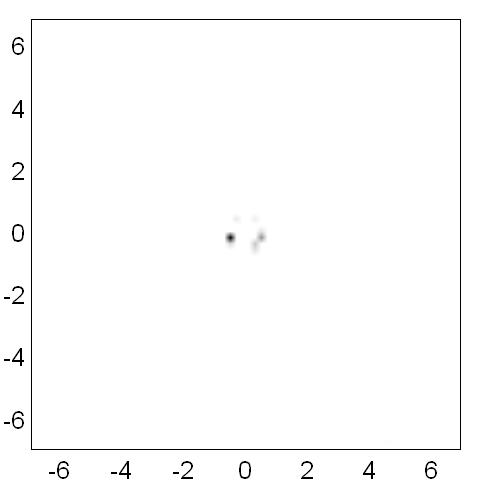} & \includegraphics[width=1.73in]{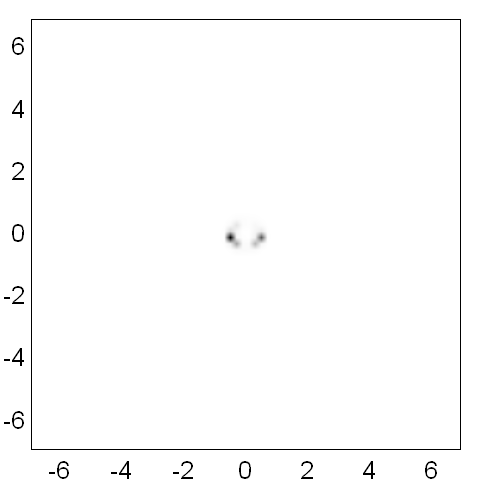} \\
		\end{tabular}
	\end{center}
	\caption[example]
   { \label{fig:sim_results_c} 
   Monte Carlo Simulation results for two equal intensity sources separated by 1$^{\circ}$. At angular separations less than the geometric resolution of the instrument, MEM reconstructs a single elongated source and CLEAN sees only a single source, while DDM and NCAR are able to fully resolve the two sources. (Axes in degrees) \\ { } \\ { }}
\end{figure}

\begin{figure}
	\begin{center}
	\begin{tabular}{@{}c@{}c@{}}
		Object Scene & Raw Image \\
		\includegraphics[width=1.73in]{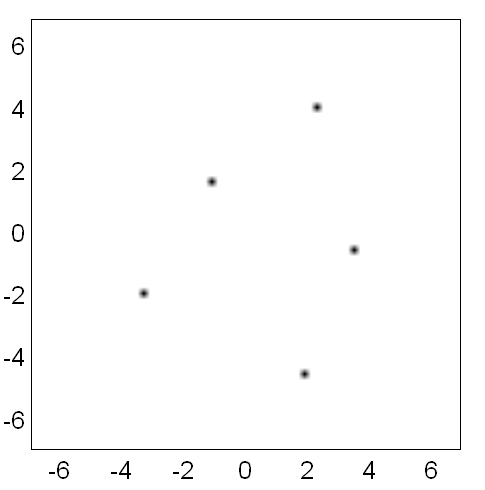} & \includegraphics[width=1.73in]{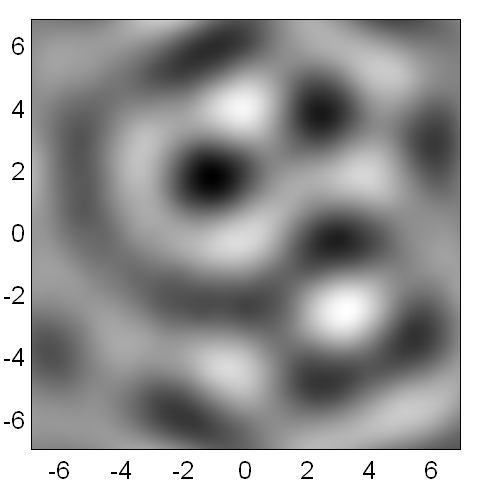} \\
		MEM & CLEAN \\
		\includegraphics[width=1.73in]{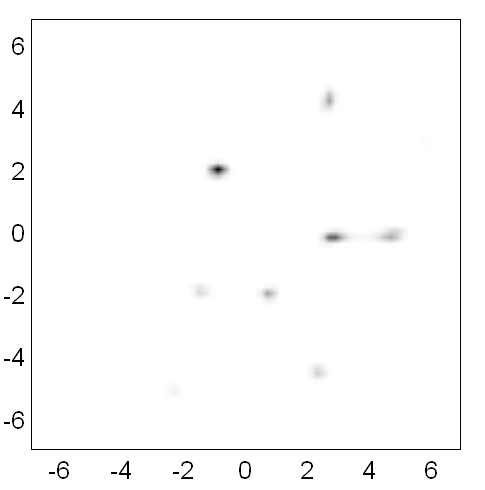} & \includegraphics[width=1.73in]{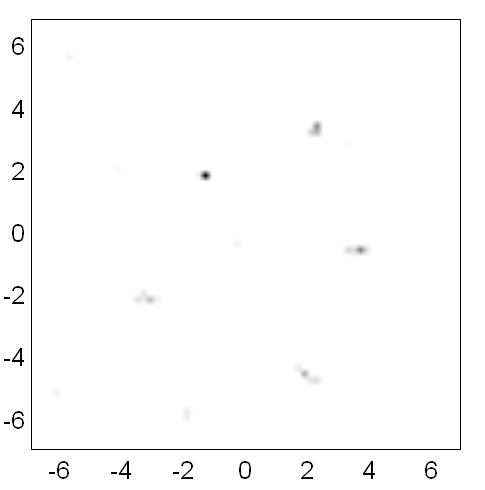} \\
		DDM & NCAR \\
		\includegraphics[width=1.73in]{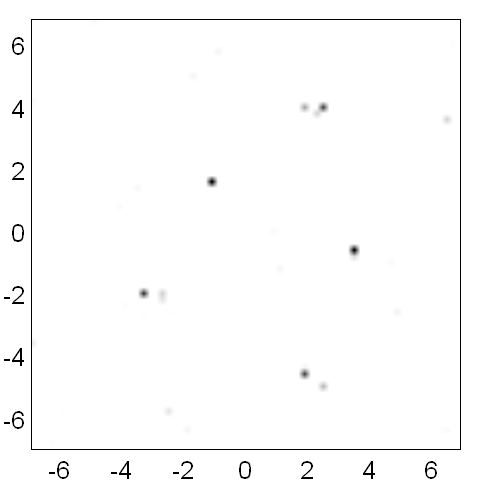} & \includegraphics[width=1.73in]{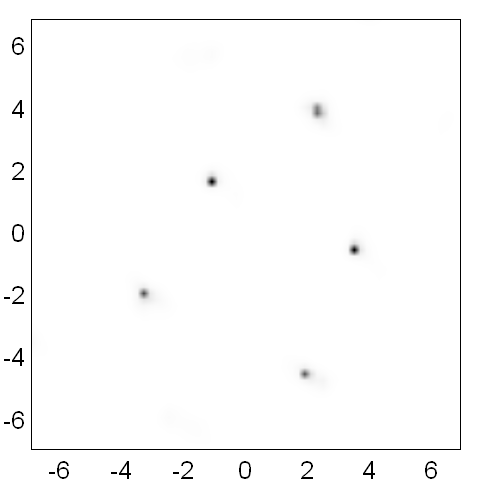} \\
		\end{tabular}
	\end{center}
	\caption[example]
   { \label{fig:sim_results_d} 
   Monte Carlo Simulation results for five equal intensity sources. Even with little background in the measurement, both MEM and DDM suffer from noise artifacts in the reconstruction due to the more complex object scene. NCAR, however, compensates for the noise and produces an image better representative of the object scene. (Axes in degrees)}
\end{figure}

\begin{figure}
	\begin{center}
	\begin{tabular}{@{}c@{}c@{}}
		Object Scene & Raw Image \\
		\includegraphics[width=1.73in]{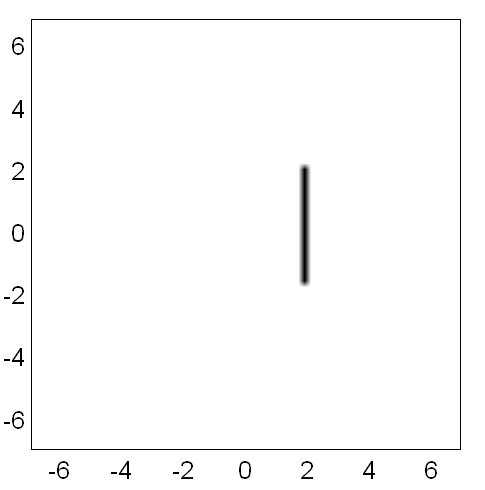} & \includegraphics[width=1.73in]{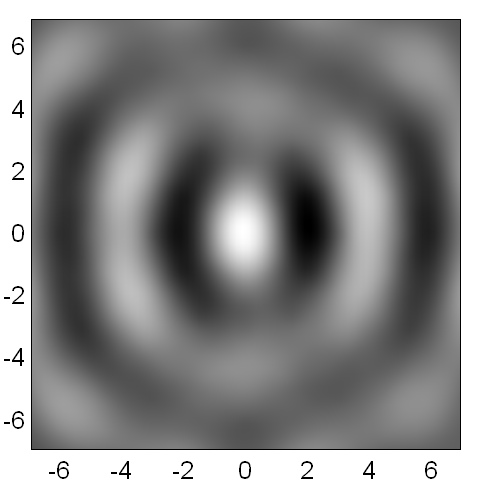} \\
		MEM & CLEAN \\
		\includegraphics[width=1.73in]{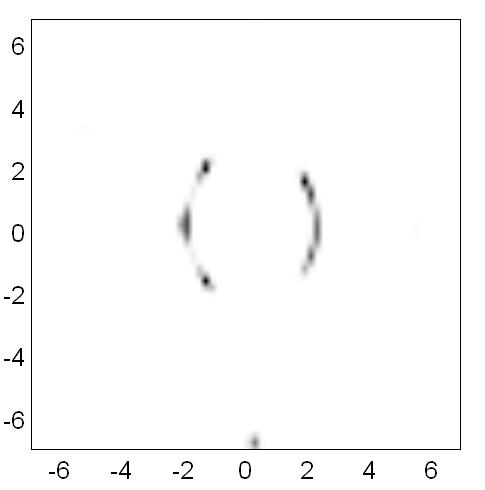} & \includegraphics[width=1.73in]{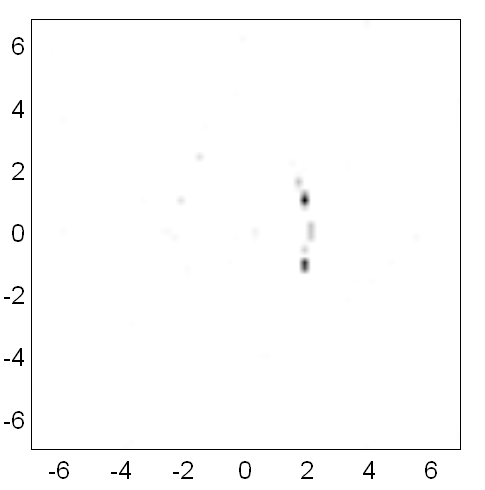} \\
		DDM & NCAR \\
		\includegraphics[width=1.73in]{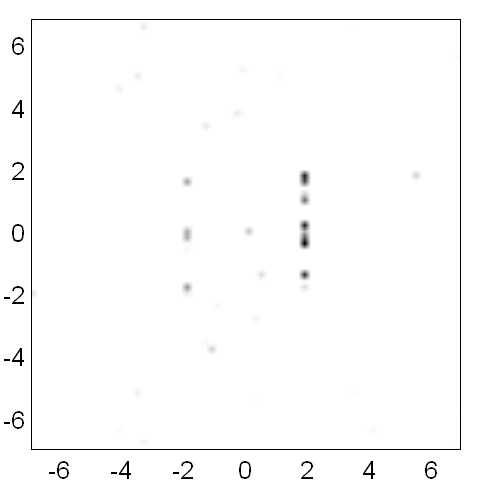} & \includegraphics[width=1.73in]{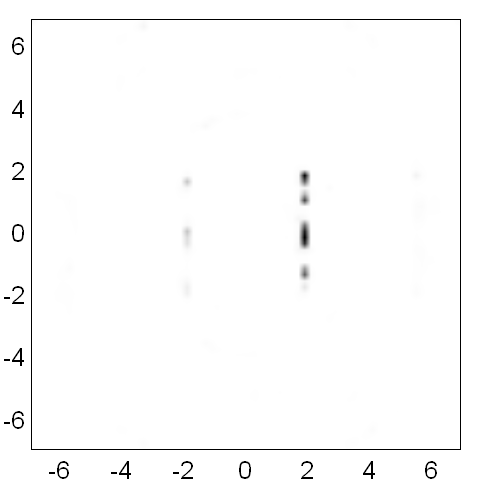} \\
		\end{tabular}
	\end{center}
	\caption[example]
   { \label{fig:sim_results_e} 
   Monte Carlo Simulation results for a line source distribution. The PSF interference patterns cause MEM to perform poorly with a complex source, while DDM continues to suffer from from noise artifacts. NCAR removes most of the spurious sources. (Axes in degrees)}
\end{figure}

\begin{figure}
	\begin{center}
	\begin{tabular}{@{}c@{}c@{}}
		Object Scene & Raw Image \\
		\includegraphics[width=1.73in]{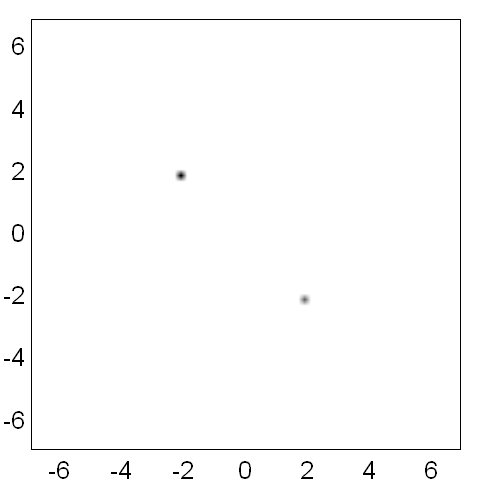} & \includegraphics[width=1.73in]{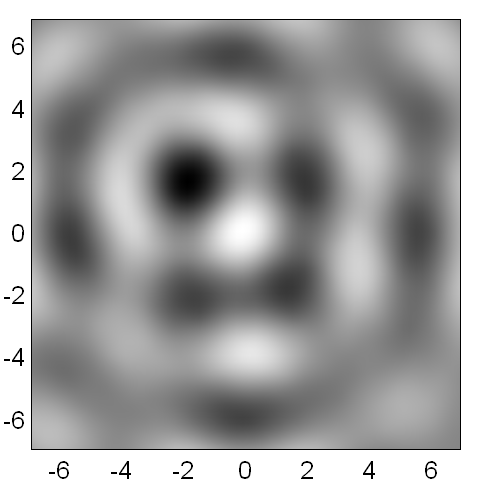} \\
		MEM & CLEAN \\
		\includegraphics[width=1.73in]{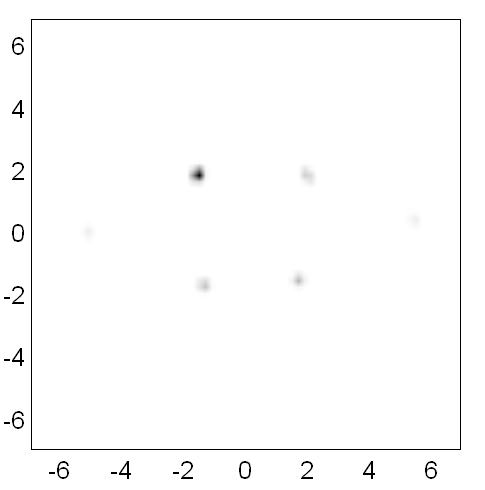} & \includegraphics[width=1.73in]{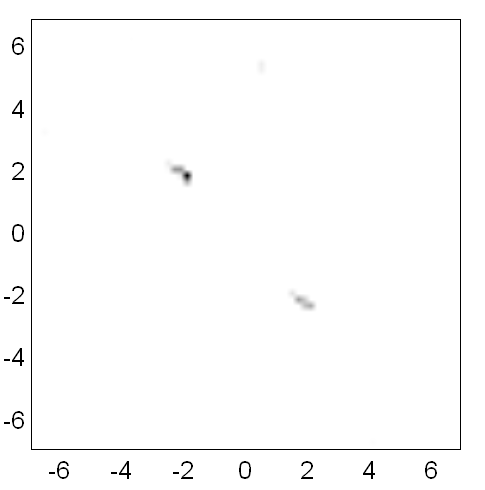} \\
		DDM & NCAR \\
		\includegraphics[width=1.73in]{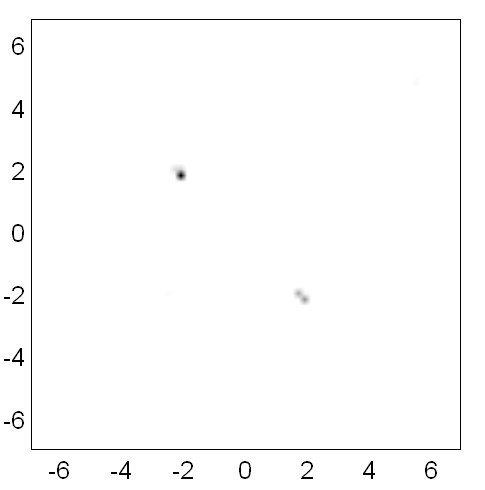} & \includegraphics[width=1.73in]{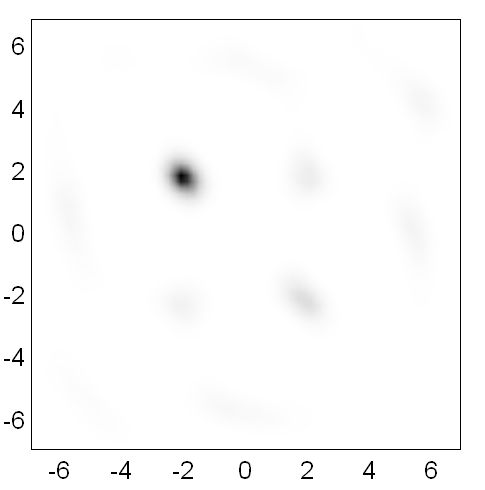} \\
		\end{tabular}
	\end{center}
	\caption[example]
   { \label{fig:sim_results_f} 
   Monte Carlo Simulation results for weak source (lower right) in the presence of a stronger source (upper left). (Axes in degrees)}
\end{figure}

In Sec. \ref{sec:results}, the NCAR technique is compared to its precursor, DDM, to demonstrate the effect of noise compensation. Additionally, we include a comparison to MEM and CLEAN, because of their wide use in astrophysical applications. For these simulations, the FOV is divided into $12\times 12$ arcmin bins, totaling 4900 pixels for a single image. Each of the results is that of a 30-minute simulated exposure. NCAR, DDM, and CLEAN are run for 10k iterations, and MEM for 50k -- 100k to assure convergence.

%Simulation
\section{Results}
\label{sec:results}

We first observed how a single point source with a measured rate of 10 Hz (over all detectors) is imaged in the presence and absence of background. When no background exists (Fig. \ref{fig:sim_results_a2}), all deconvolution techniques reconstruct the scene reasonably well, although spurious point sources are faintly visible in the MEM reconstruction.

When a background rate of 500 Hz is included (Fig. \ref{fig:sim_results_a}), the MEM, CLEAN, and DDM reconstructions exhibit noise fluctuations that are manifest as spurious peaks in the image. The true source itself, however, remains relatively unaffected. The NCAR image shows the presence of diffuse background features, and the true source peak has broadened. Due to the nature of NCAR, the width of a point source reconstruction narrows as the SNR grows. This can be a desirable feature of an image, where a ``blur'' is typically considered an uncertainty on source location. As we continue to show, this blurring characteristic does a good job of removing spurious noise fluctuations.

In Figs. \ref{fig:sim_results_b}-\ref{fig:sim_results_c}, object scenes with two sources, each with a measured rate of 50 Hz, and a background contribution of 100 Hz, are simulated. The background is kept relatively low here to demonstrate the intrinsic ability of NCAR to achieve super-resolution. In Fig. \ref{fig:sim_results_b}, the sources are separated by 3$^{\circ}$, greater than the 1.9$^{\circ}$ geometric resolution of the instrument. NCAR produces the result that is the most free of spurious peaks and other degrading artifacts. Due to the broadened nature of the RM PSF, CLEAN in particular shows difficulty in resolving sources at this separation. In Fig. \ref{fig:sim_results_c}, the sources are moved to a separation of 1$^{\circ}$, about half the intrinsic resolution of the instrument. MEM reconstructs an elongated structure, while CLEAN reconstructs only a single point source. Additional simulations have confirmed that MEM and CLEAN are both limited by the geometric resolution of the instrument. DDM and NCAR resolve both sources and provide the desired super-resolution. This result has also been previously demonstrated with measured laboratory data, resulting in a resolving power of 20$'$, or $\sim 8 \times$ the geometric resolution of the instrument \cite{BuddenIEEE2009}. Again, NCAR produces the most accurate reconstruction of the object scene.

NCAR also provides image reconstructions for more complicated source distributions.  Fig. \ref{fig:sim_results_d} shows the results for 5 sources, each with a measured source rate of 50 Hz and equivalent background rate. MEM is capable of resolving some, but not all of the sources entirely, while CLEAN performs slightly better. DDM suffers from spurious peaks near the true sources. NCAR, however, reconstructs each of the sources well, with no visible noise, but instead, a slight blurring.

A line source distribution (Fig. \ref{fig:sim_results_e}) is also simulated. To examine the ability of the reconstruction techniques to resolve an extended source, the background rate is set to zero, and the line source has a total measured rate of 4.5 kHz. MEM performs poorly with a significant spurious reconstruction to the left due to the interference of sidelobes. The CLEAN result has no significant spurious reconstructions, but the line is poorly resolved. DDM suffers from some spurious peaks arising from noise and interfering PSFs from the complex nature of the object scenes. NCAR provides a reconstruction of the line similar to DDM with most of the spurious peaks smoothed out.

The ``broken'' lines observed in the DDM and NCAR reconstructions in Fig. \ref{fig:sim_results_e} are not due to sensitivity variations. As seen in Fig. \ref{fig:labrat_sens}, the sensitivity is relatively uniform, and the low-sensitivity sky regions are uncorrelated with the line breaks. Rather, this is a consequence of the statistical noise in the image causing the deconvolution to converge on an imperfect solution to the data. This effect is seen to increase for more complex object scenes (i.e. the number of non-trivial sky bin values increases), and is an inherent property of the RM response, not the deconvolution algorithm. The effect could be reduced by increasing the exposure time (thus improving SNR).

In order to observe two weak sources imaged simultaneously, a 25 and 15 Hz source, in the presence of 500 Hz background, are simulated in Fig. \ref{fig:sim_results_f}. MEM performs poorly, showing spurious reconstructions indistinguishable from the weaker of the two sources. CLEAN and DDM perform the best, with two sources clearly reconstructed with relative fluxes apparent. The ``blurring'' characteristic of NCAR, however, means that the weaker of the two sources is blurred to a greater degree, suppressing its apparent brightness further. For higher SNR, this discrepancy disappears.

% Discussion
\section{Discussion}

For imaging modalities with 50\% transmission, the coded aperture offers the maximum sensitivity possible. By moving to the temporal domain for photon modulation, the RM gains a unique advantage in that the measurement vector is effectively a continuous function, and so finer sub-sampling allows one to achieve super-resolution with an appropriate reconstruction algorithm. The RM does, however, suffer in sensitivity due to its non-ideal response.

At the geometric angular resolution $\delta\theta = a/L$ (i.e., with super-resolution factor $\eta=1$), a coded aperture has an intrinsic sensitivity advantage over an RM by a factor $0.50/0.31 \approx 1.6$ (Eqs. \ref{eq:SNR}, \ref{eq:SNR_RM}). The width/diameter $b$ of the detector pixel is a fraction $\alpha$ of the mask element size $a$: $b = a/\alpha$ (typically, $\alpha=1$ for an RM and 2-4 for a coded aperture). For good efficiency at high energies, the detector width and thickness should be at least equal to the photon interaction length, $1/\sigma\rho$, implying that
\begin{equation}
	\sigma\rho \gtrsim \frac{\alpha}{L \, \delta\theta};
\end{equation}
i.e., there is a minimum $\sigma\rho$ (or equivalently a maximum energy) at which a particular angular resolution is achievable. For example, at 511 keV, with $L=1$ m and $\alpha = 2$, the angular resolution of a coded aperture imager is limited to roughly $\delta\theta \gtrsim \alpha / \sigma\rho L \approx 2.6^{\circ}$. In other words, at hard X-ray energies and above, a coded aperture has a sensitivity advantage over an RM but cannot take advantage of its full potential for excellent angular resolution. At high energies, the super-resolution capability of an RM with NCAR allows it to provide good angular resolution although there is a penalty of $1.6 \eta^2$ in the sensitivity compared to the coded aperture. At lower energies, the RM can utilize its super-resolution capability, while the coded aperture is restricted to the geometric angular resolution.

%The individual application and imaging requirements will dictate which of these modalities is favored. When good sensitivity and fidelity are desired, the coded aperture is an attractive solution. When sensitivity is less important than instrument cost, or resolving power beyond the instrinsic limits of the instrument are subsequently desired, the RM and RMC weigh favorably. At high energies, the RM is favored over the RMC for more uniform sensitivity curve and its wider achieveable FOV.

%Without the ability to resolve beyond the instrinsic limit of the instrument, the temporal modulators offer no advantage. As demonstrated with simulations of an RM with the NCAR algorithm, super-resolution is achievable. However, this ability comes with a price; for each factor of $\eta$ of enhanced resolution, the object scene is subdivided into an additional $\eta^2$ bins. Individual source strength must therefore increase by a factor of $\eta^2$ to maintain an equivalent statistical significance for each sky bin observation. Thus, the instrument becomes less efficient by $\eta^{-2}$ for every factor $\eta$ of increased resolution. Where the use of an RM becomes most attractive, however, is in the high-energy, wide-FOV regime, particularly for survey missions. In this case, the RMC exhibits auto-collimation due to a thick grid. Since an RM is capable of spacing the slats to a greater pitch, while still achieving the same angular resolution (albeit with less efficiency), the maximum observable source angles are increased. 

When compared to the RMC, the RM maintains a sensitivity advantage of $0.31/0.15 \approx 2$, due to its higher mask transmission. Additionally, the moderate spatial resolution of the RM detection plane increases the number of independent measurements of the object scene, contributing to enhanced fidelity of the reconstructed image and smoothing the sensitivity curve across the sky, and the absence of a thick secondary grid reduces instrument weight.

An algebraic technique such as NCAR is necessary to exploit the super-resolution capability of the RM while simultaneously suppressing noise fluctuations. NCAR will not, however, be suitable for all scenarios. As shown, its weakest performance is in resolving two weak sources with different measured rates. Because of the inherently low SNR, the peaks are ``blurred'' to different degrees, suppressing the weaker of the two and thus not accurately depicting relative source strength. In the other cases presented, however, NCAR performs quite well, reconstructing the object scene and suppressing noise fluctuations that typically plague algebraic solutions. In scenarios where the size of the source is of importance, NCAR will not produce good results with low SNR. If the ultimate desire is to survey and locate true sources, however, NCAR generates images free of spurious reconstructions with a visual representation of the locational uncertainty of a particular source measurement.

% Conclusion
\section{Conclusion}

A rotating modulator is an instrument capable of imaging at hard x-ray and gamma-ray energies. The instrument is composed of a single grid of slats above a collection of circular non-imaging detector elements. While the instrument response for the RM is non-ideal (the sensitivity is approximately 62\% of that of an equivalent coded aperture), the near-continuous nature of the observed count profiles enables an analysis which can go beyond the instrinsic resolution as defined by the geometry, whereas the coded aperture cannot. When compared to the RMC, the RM is able to achieve higher and more uniform sensitivity, does not suffer from mechanical collimation at high energies, and has lower overall weight.

We have found that algebraic solutions are the only reconstruction techniques that can both achieve super-resolution and perform relatively fast. Algebraic methods, however, have in the past suffered from poor reconstructions due to the requirement of strict agreement with the noisy data. We have developed a novel technique, NCAR, based on an algebraic solution and non-linear physical constraints. A noise-compensation factor derived directly from the Poisson uncertainty in the data is added to a Gauss-Seidel iteration. By performing a running average on the successive results after a specific agreement criterion has been met, the spurious peaks arising from noise in the data are effectively suppressed, while the true sources remain. The uncertainty of the true source location due to noise in the data is manifest as a blurring of the scene rather than spurious peaks elsewhere in the image.

When compared to other common deconvolution algorithms (MEM, CLEAN, and DDM), NCAR provides higher fidelity in most cases. Furthermore, the technique is general enough that it could be used in imaging and other applications where deconvolution is required in the presence of large background.

% Acknowledgement
\section*{Acknowledgment}
This work has been supported in part by US DOE NNSA Cooperative Agreement DE-FC52-04-NA25683.

B. Budden thanks the Louisiana Board of Regents under agreement NASA/LEQSF(2005-2010)-LaSPACE and NASA/LaSPACE under grant NNG05GH22H; and the Coates Foundation at LSU for support during this project.

We wish to thank the anonymous reviewers for their comments and suggestions. Additionally, we thank K. Matthews for useful discussions.

% Can use something like this to put references on a page
% by themselves when using endfloat and the captionsoff option.
\ifCLASSOPTIONcaptionsoff
  \newpage
\fi

% Bibliography
\bibliographystyle{IEEEtran}
% argument is your BibTeX string definitions and bibliography database(s)
\bibliography{BIBLIOGRAPHY}

%\begin{IEEEbiography}{Brent Budden}
%Biography text here.
%\end{IEEEbiography}

% if you will not have a photo at all:
%\begin{IEEEbiographynophoto}{Brent Budden}
%Biography text here.
%\end{IEEEbiographynophoto}
%
%\begin{IEEEbiographynophoto}{Gary Case}
%Biography text here.
%\end{IEEEbiographynophoto}
%
%\begin{IEEEbiographynophoto}{Michael Cherry}
%Biography text here.
%\end{IEEEbiographynophoto}

\end{document}